\documentclass[twocolumn,aps,prb,showpacs,floatfix,epsfig]{revtex4}

\usepackage{graphicx}%
\usepackage{dcolumn}
\usepackage{bm}%

\newcommand \be{\begin{equation}}
\newcommand \ee{\end{equation}}
\newcommand \ba{\begin{eqnarray}}
\newcommand \ea{\end{eqnarray}}
\begin{document}
\title
{\Large\bf
Theory of traveling filaments in bistable semiconductor
structures}

\author{\bf Pavel Rodin \cite{EMAIL}}
\affiliation
{Institut f{\"u}r Theoretische Physik, Technische Universit{\"a}t
Berlin,
Hardenbergstrasse 36, D-10623, Berlin, Germany\\ and \\
Ioffe Physicotechnical Institute of Russian Academy of Sciences,
Politechnicheskaya 26, 194021, St. Petersburg, Russia
}

\setcounter{page}{1}
\date{\today}

\hyphenation{cha-rac-te-ris-tics}
\hyphenation{se-mi-con-duc-tor}
\hyphenation{fluc-tua-tion}
\hyphenation{fi-la-men-ta-tion}
\hyphenation{self--con-sis-tent}
\hyphenation{cor-res-pon-ding}
\hyphenation{con-duc-ti-vi-ti-tes}


\begin{abstract}
We present a generic nonlinear model for current filamentation in
semiconductor structures with S-shaped current-voltage
characteristics. The model accounts for Joule self-heating of a
current density filament. It is shown that the self-heating leads to
a bifurcation from static to traveling filament. Filaments start to
travel when increase of the lattice temperature has negative impact on
the cathode-anode transport. Since the impact ionization rate
decreases with temperature, this occurs for  a wide class of
semiconductor systems whose bistability is due to the avalanche impact
ionization. We develop an analytical theory of traveling filaments
which reveals the mechanism of filament motion, find the condition
for bifurcation to traveling filament, and determine the filament velocity.

\end{abstract}


\maketitle





\section{Introduction}

Bistable semiconductor systems with S-shaped current-voltage
characteristics exhibit spontaneous formation of current density
filaments --- high current density domains in the low current
density environment.\cite{RID63,VOL69,BONCH75,SZE,SCH87,SHAW}
Current filamentation typically develops due
to the spatial long wavelength instability
--- known as Ridley instability \cite{RID63} --- of the uniform state
with negative differential conductance when the device is operated
via sufficiently large load resistance, \cite{WAC95,ALE98} Fig.~1.
Formation of current filaments potentially leads to thermal
destruction of semiconductor structure and thus is an
important scenario of semiconductor device failure.\cite{SZE,SHAW}
Current filaments  have been studied in bulk
semiconductors,\cite{VOL69,SCH87,VOL67,BASS70} thin semiconductor
films,\cite{PRETTL95,PRETTL98,PRETTL98a,PRETTL00,SCHWARZ00} and
layered structures of semiconductor
devices.\cite{OSI70,OSI71,OSI73,JAG86,GOR92,GOR02}
Over last decade, the focus has been
shifted from static filaments to complex spatio-temporal dynamics of
current density patterns.\cite{KER94,NIE95,AOK00,SCH01}
Modern experimental techniques based
on electron scanning microscopy,\cite{WIE95}
detection of infrared radiation,\cite{NIE92} and interferometric
mapping\cite{POG02} provide means for  direct
observation of current density dynamics during filamentation.
In regard to pattern formation and nonlinear phenomena,
bistable semiconductors have much in common with other spatially
distributed active media such as gas discharge systems,
chemical and biological systems.\cite{KER94,HOH93,KUR88,MIK94,BLUE,BUSS98}

Apart from the well understood bifurcation which leads to temporal
periodic or chaotic oscillations,\cite{WAC94,NIE96a,BOS00,PLE02}
static filament may undergo a secondary bifurcation that
leads to traveling filament. Lateral movement of filaments along the
device has been observed experimentally for different types of
semiconductor structures.\cite{NIE92,POG02,PEN88,POG03} Remarkably, in
all these structures the S-shaped characteristic is associated with
avalanche impact ionization. Since impact ionization coefficients
decrease with temperature,\cite{SZE}
the onset of filament
motion has been attributed to the self-heating of the filament:
\cite{WAC91} the filament is expected to migrate to a colder region
as temperature at its initial location increases. When current filamentation
is unavoidable, the migration is
desirable in applications because the filament motion delocalizes
the heating of the semiconductor structure and thus reduces the hazard of thermal
destruction.

The purpose of this paper is to describe the mechanism of  filament
motion, to develop a generic nonlinear model for
traveling filament,
to find the condition for bifurcation to traveling filament, and to
determine the filament velocity. We demonstrate that the transition from
static to traveling filament due to the Joule self-heating represents a
generic effect which potentially appears in any bistable
semiconductor structure
provided that transport in the cathode-anode direction is sensitive
to temperature.

\section{Model of a bistable semiconductor system}

For many semiconductors and semiconductor devices
current filamentation can be described by
a two-component reaction-diffusion model
which consists of a
partial differential equation for the bistable element
and an integro-differential Kirchhoff's equation for the external
circuit \cite{WAC95,ALE98,MEI00}
\begin{eqnarray}
\label{RD0}
\frac{\partial a}{\partial t} = \nabla_{\perp}(D_a(a) \nabla_{\perp} a) +
f(a,u),\;\nabla_{\perp} \equiv  {\bf e}_x \partial_x + {\bf e}_y \partial_y,  \\
\label{GC0}
\tau_u \frac{d u}{d t} = U_0 - u - R \int_S J(a,u) dx dy ,\; \tau_u
\equiv RC.
\end{eqnarray}
Here the variable $a(x,y,t)$ characterizes the internal state of the
device and the variable  $u(t)$ is the voltage over the device,
$U_0$ is the total applied voltage, $J$ is the current density,
$R$ is the load resistance connected
in series with the device, $C$ is the effective capacitance of the
device and the external circuit, and $S$ is the device cross-section
(Fig.~1). The local kinetic function $f(a,u)$ and the $J(a,u)$ dependence
contain all information about transport in the
cathode-anode (vertical) direction; the diffusion coefficient $D_a (a)$
characterizes lateral coupling in the spatially extended element.
In the bistability range $u_{\rm h} < u < u_{\rm th}$ (see Fig.~1) the local
kinetic
function $f(a,u)$ has three zeros $a_{\rm off},a_{\rm int},a_{\rm on}$
corresponding to off, intermediate, and on branches of current voltage
characteristic. It has one zero outside the bistability range.
For the homogeneous state, the local dependence $a(u)$ is calculated
from $f(a,u)=0$, and then inserted into $J(a,u)$ in order to determine
the local current-voltage characteristic $J(u) \equiv J(a(u),u)$.
Typically $\partial_u f, \partial_a j, \partial_u j > 0$. The model
(\ref{RD0}),\ (\ref{GC0}) belongs to a class of activator-inhibitor
models with global inhibition.\cite{SCH01} Variables $a$ and $u$
serve as activator and inhibitor, respectively. The time scale of the
activator $a$ is $\tau_a = \partial_a f^{-1}$.\cite{WAC95} Specific
functional forms $f(a,u)$, $j(a,u)$, and $D_a(a)$ have been
derived for various semiconductors and semiconductor
structures.\cite{WAC95,ALE98,VOL67,VOL69,BASS70,OSI71,OSI73,GOR92,WAC94,MEI00,ROD03}
The physical meaning of $a$ depends on the particular type of
bistable structure: $a$ corresponds to the bias of the emitter
$p\,$-$n$ junction for  avalanche transistors,\cite{OSI73} thyristors
\cite{GOR92} and thyristorlike structures,\cite{GOR02}  interface
charge for heterostructure hot electron diode,\cite{WAC94} electron
charge stored in the quantum well for bistable double barrier
resonant tunneling diode,\cite{MEI00,ROD03} {\it etc.}

The two-component model (\ref{RD0}),\ (\ref{GC0})
is capable to describe steady filaments\cite{ALE98} as well as
temporal oscillation of current density filaments.\cite{WAC94,BOS00,PLE02}
However, its solutions do not include traveling filaments.

The self-heating of the filament has an impact on filament dynamics
when vertical transport is sensitive to temperature. In terms
of Eq.\ (\ref{RD0}) it means that the local kinetic function $f$
depends on the lattice temperature $T$.
We get $\partial_T f < 0$ when temperature suppresses the vertical
transport. Since the thermal
diffusion length $\ell_T$ is typically much larger than the device width $w$,
the heat dynamics
in the device can be modelled by the two-dimensional equation
\begin{eqnarray}
\label{heat1}
c \rho w \; \frac {\partial T}{\partial t}
= \kappa w \; \Delta_{\perp} T + J u - \gamma(T &-& T_{\rm ext}), \\ \nonumber
& &\Delta_{\perp} \equiv  \partial_x^2 + \partial_y^2.
\end{eqnarray}
Here $T$ corresponds to the mean value of temperature along the
device vertical direction, $c$, $\rho$, and $\kappa$ are specific
heat, density, and heat conductivity of the semiconductor material,
respectively. The second and the third terms on the right-hand side of
Eq.\ (\ref{heat1}) describe Joule heating and cooling due to 
contact with environment, respectively. $T_{\rm ext}$ is the temperature
of the external cooling reservoir. For multilayer structures with
classical transport, $T_{\rm ext}$ is usually a room temperature.
The coefficient $\gamma$ is a heat transfer coefficient (per unit square
of the structure)  that characterizes the efficiency of external
cooling. The characteristic relaxation time $\tau_T$, diffusion
length $\ell_T $, and propagation velocity $v_T$ are given by
\begin{equation}
\tau_T \equiv \frac{c \rho w}{\gamma},  \;
\ell_T \equiv \sqrt{\frac{\kappa w}{\gamma}}, \;
v_T \equiv \frac{\ell_T}{\tau_T} = \sqrt{\frac{\kappa \, \gamma}{c^2
\rho^2 w}}.
\label{thermal_scales}
\end{equation}

The modified model is given by the following set of equations:
\begin{eqnarray}
&&\frac{\partial a}{\partial t} =
\nabla_{\perp} (D_a(a) \nabla_{\perp} a) + f(a,u,T), \label{RD1} \\
&&\tau_T \, \frac{\partial T}{\partial t} = \ell_T^2 \Delta_{\perp} T +
\bigl( {J u}/{\gamma} +
T_{\rm ext} - T \bigr),
\label{RD_T1}\\
&&\tau_u \frac{d u}{d t} = U_0 - u - R \int_S J(a,u,T) \, dx dy \label{GC1}.
\end{eqnarray}
The variable $T$ plays a role of the second inhibitor.

In the following we assume that the device is elongated along the $x$
direction ($L_x \gg L_y$) and take only this lateral dimension into account.
We also neglect direct effect of temperature on the current density
$J$ in Eq.\ (\ref{GC1}).

\section{Current filament and mechanism of its motion}

Current density filament in a long structure represents a
domain of the high current density state embedded into a
low current density state (Fig.~2). The width
of the filament wall $\ell_f$ is of the order of
$\ell_f \sim \sqrt{D_a / \partial_a f}$.\cite{WAC95} For $T = T_{\rm ext}$
the voltage $u_{\rm co}$
over the device with steady filament is chosen by the condition
known as ``equal area rule"  \cite{VOL69,SCH87}
\begin{equation}
\int_{a_{\rm off}}^{a_{\rm on}} f(a,u_{\rm co},T = T_{\rm ext}) D_a(a) da = 0,
\label{EAR}
\end{equation}
which ensures that production and annihilation of the inhibitor $a$
in the filament wall compensate each other. When the integral
in (\ref{EAR}) is positive or negative, the balance is broken,
and the filament walls move in such a
way that the high current density state or the low current density
state, respectively, expand. Since
$\partial_u f > 0$, the filament expands for $u > u_{\rm co}$ and shrinks
for $u < u_{\rm co}$. The filament has neutral stability with respect
to the lateral shift when $T$ is kept constant.\cite{ALE98}

The current-voltage characteristic of
the filament in long structure is practically
vertical \cite{VOL69,SCH87,ALE98,MEI00} and bends in the upper and lower
parts when the filament becomes narrow (Fig.~1). The bent parts correspond to
current density intervals
$[J_{\rm off}; J_{\rm off}+(\ell_f/L_x)J_{\rm on}]$
and $[J_{\rm on}-(\ell_f/L_x)J_{\rm on}; J_{\rm on}]$,
where $J_{\rm on}$ and $J_{\rm off}$ are current densities in the
high current density state and the low current density state, respectively.
These intervals are  negligible
for $L_x \gg \ell_f$.\cite{VOL69,ALE98,MEI00}

The current filament is stable for sufficiently large load resistance
$R$ and small relaxation time $\tau_u$.\cite{BASS70,ALE98}
We assume that $\tau_u \ll \tau_a, \tau_T$, and hence Eq.\ (\ref{GC1})
essentially represents a constraint
imposed on the dynamics determined by Eqs.\ (\ref{RD1}),\ (\ref{RD_T1}).
Without loss of generality we can also assume that the regime of the
external circuit is close to the  current-controlled regime:
$U_0 \gg u_{\rm co}$ and the total current
$I \approx U_0 / R$ is constant.
The width of the filament is directly proportional to the total
current $I$
\begin{equation}
W = \frac{1}{L_y} \frac{I - L_x L_y \cdot J_{\rm off}}{J_{\rm on} -
J_{\rm off}} \approx
\frac{1}{L_y} \frac{I}{J_{\rm on}},
\label{width_current}
\end{equation}
where the last equality takes into account that typically
$J_{\rm on} \gg J_{\rm off}$.

Qualitatively, the mechanism of the filament motions in presence of
self-heating is the following. With increase of temperature the
stationary balance (\ref{EAR}) within the filament wall is broken due
to the temperature dependence of the local kinetic function
$f(a,u,T)$. As far as the temperature profile $T(x)$ is symmetric,
the left and right walls of the filament are equal and filament would
either expand or shrink. This is forbidden since the total current is
conserved by the global constraint (\ref{GC1}). Increase of
temperature is compensated by deviation of $u$ from $u_{\rm co}$ in
such a way that the stationary balance within the filament wall is
restored. Since $\partial_u f > 0$ and $\partial_T f < 0$, the
voltage increases, but the  filament stays steady. In contrast, for
antisymmetric temperature fluctuation the balance is disturbed
differently in the left and right filament walls, becoming positive
at one side and negative at another side. Potentially, this
spontaneous instability leads to the motion of a filament as a whole
preserving the total current. In the traveling filament the
temperature at the back edge exceeds the temperature at the leading
edge due to the heat inertia of the semiconductor structure. Hence
the filament motion becomes self-sustained.

Below we present an analytical theory of this
effect. The theory is based on the following assumptions:

(i)  the effect of self-heating is small and  can be considered as a
perturbation, hence the local kinetic function can be linearized
near $T = T_{\rm ext}$  and $u = u_{\rm co}$ as
\begin{eqnarray}
\label{expansion}
f(a,u,T) &=& f(a,u_{\rm co}, T = T_{\rm ext})
\\ \nonumber
&+&(u - u_{\rm co}) \, \partial_u f +
(T - T_{\rm ext})\, \partial_T f  ;
\end{eqnarray}

(ii) the transverse dimension of the semiconductor structure is
large in the sense that  $L_x \gg \ell_f, \ell_T$, therefore
only wide filaments with $W \gg \ell_f$ are relevant and the effect
of boundaries is negligible;

(iii) the width of the filament wall $\ell_f$ is much smaller than
the thermal diffusion length $\ell_T$ and therefore the temperature
variation within the filament walls can be neglected.

\section{Stationary motion of a filament: general properties}

\subsection{Model equations in the co-moving frame}

For stationary motion of a filament with a constant velocity $v$
the solution of Eqs.\, (\ref{RD1}),\ (\ref{GC1}),\ (\ref{RD_T1})
has a form:
\begin{equation}
\nonumber
a(x,t) = a(x - vt), \;
T(x,t) = T(x - vt), \;
u = {\rm const}.
\end{equation}
In the co-moving frame
$\xi = x - vt$, Eqs.\ (\ref{RD1}),\ (\ref{GC1}),\ (\ref{RD_T1})
are
\begin{eqnarray}
&&(D_a(a) a^{\prime})^ {\prime} + v \, a^{\prime} + f(a,u,T) = 0,
\label{c_m_1}
\\ \label{c_m_2}
&&\ell_T^2 \, T^{\prime \prime} + v \, \tau_T \, T^{\prime}
 + \bigl( {J u}/{\gamma} + T_{\rm ext} - T \bigr) = 0,
\\
&&U_0 - u - R \, L_y \, \langle J(a,u) \rangle = 0.
\label{c_m_3}
\end{eqnarray}
Here the prime $(...)^{\prime}$ denotes the derivative with respect to $\xi$
and angular brackets $\langle ...  \rangle$ denote integration
over $\xi$.
For the sufficiently large $L_x$
the boundary conditions are given by
\begin{equation}
a(\xi) \rightarrow a_{\rm off}(u), \; T(\xi) \rightarrow T_{\star} \qquad
{\rm for} \qquad \xi \rightarrow \pm \infty,
\label{BC}
\end{equation}
where
\begin{equation}
\label{T_low_star}
T_\star \equiv T_{\rm ext} + { J_{\rm off}(u_{\rm co})u_{\rm co}}/{\gamma}
\end{equation}
is the stationary temperature corresponding to the low current
density state.

\subsection{Filament velocity and voltage over the structure}

In order to calculate the velocity of the filament and
the voltage on the structure with traveling filament, we
note that for given $u$ and $T$ both filament walls
can be treated as propagating fronts in bistable medium
and use the standard formula \cite{MIK94,SCH01,MEI00}
\begin{equation}
\label{standard}
v = \frac{2 \int_{a_{\rm off}}^{a_{\rm on}} f(a,u,T)\, D_a (a) \, da}
{\langle D_a(a) (a_0^{\prime})^2 \rangle},
\end{equation}
where the filament profile is approximated by the stationary profile
$a_0 (x)$. Eq.\ (\ref{standard}) implies that the velocity $v$ is
proportional to the disbalance of production and annihilation of the
activator $a$ in the filament wall. Here $T$ is the temperature
within the filament wall under consideration, which is taken as
constant according to the assumption $\ell_f \ll \ell_T$. The factor
2 in the nominator appears because $a_0$ corresponds to the pattern
that consists of two fronts. Positive and negative velocities
correspond to the propagation of the high current density state into
the low current density state and {\it vice versa}, respectively.

Linearizing the local kinetic function
according to Eq.\ (\ref{expansion})
and taking into account Eq.\ (\ref{EAR}), we obtain
\begin{eqnarray}
\label{vel_auxiliary_1}
&& v(u,T) = \frac{2}{\langle D_a(a) (a_0^{\prime})^2 \rangle} \\
&& \times \int_{a_{\rm off}}^{a_{\rm on}}
\bigl[(u-u_{\rm co}) \, \partial_u f + (T - T_{\rm ext}) \,
\partial_T f  \bigr]
\, D_a(a) \, da.
\nonumber
\end{eqnarray}
Hereafter, all derivatives are taken at $u = u_{\rm co}$ and
$T = T_{\rm ext}$.

Equation (\ref{vel_auxiliary_1}) should be applied separately to the
left and the right filament walls. In terms of Eq.\
(\ref{vel_auxiliary_1}) the respective velocities have the same
absolute value but different signs
\begin{equation}
v(u,T_R) = - v(u, T_L),
\label{vel_auxiliary_2}
\end{equation}
where $T_R$ and $T_L$ are temperatures at the right and the left
walls, respectively. Equations
(\ref{vel_auxiliary_1}) and (\ref{vel_auxiliary_2}) yield together
\begin{eqnarray}
&&v = \frac{T_R - T_L}{{\langle D_a(a)\, (a_0^{\prime})^2 \rangle}}
\int_{a_{\rm off}}^{a_{\rm on}} \partial_T f \,D_a(a) \, da,
\label{velocity_3} \\
&&u = u_{\rm co} +  B \, \left( \frac{T_L + T_R}{2} - T_{\rm ext}
\right),
\label{voltage}\\
&& B \equiv
- \frac{\int_{a_{\rm off}}^{a_{\rm on}} \partial_T f \,D_a(a)\, da}
{\int_{a_{\rm off}}^{a_{\rm on}} \partial_u f \,D_a(a)\, da}.
\nonumber
\end{eqnarray}
According to Eqs.\ (\ref{velocity_3}),\ (\ref{voltage}) the filament velocity
is proportional to the difference of temperatures at the filament edges
($T_L - T_R$), while the voltage deviation from $u_{\rm co}$ is
proportional to the mean value $(T_L+T_R)/2$. For $\partial_T
f < 0$ the coefficient $B$ is positive, and hence the voltage
increases with the increase of temperature. The filament moves to the
right ($v > 0$) when $T_L > T_R$ and to the left ($v < 0$) when $T_L
< T_R$.

It is worth to mention that formula (\ref{velocity_3}) derived for
the wide filament with narrow walls
is a special case of the general expression
\begin{equation}
v = - \frac{\langle \partial_T f \, D_a(a) \, T(\xi)\, a_0^{\prime}  \rangle}
{\langle D_a(a) (a_0^{\prime})^2 \rangle},
\label{velocity_2}
\end{equation}
which is applicable for any filament shape and temperature profile
$T(\xi)$ satisfying the boundary condition (\ref{BC}). Similar to
Eq.\ (\ref{standard}), equation (\ref{velocity_2}) can be obtained by
multiplying  Eq.\ (\ref{c_m_1}) by $D_a (a) a^{\prime}$, integrating
over $\xi$ and using the expansion (\ref{expansion}).

Eqs.\ (\ref{velocity_3}),\ (\ref{voltage}) do not refer exclusively to the
case of Joule self-heating. They are applicable regardless
of the origin of the inhomogeneous temperature profile in the
semiconductor structure. We conclude that when temperature suppresses
the vertical transport ($\partial_T f < 0$), as it happens in case of
impact ionization mechanism of S-type characteristic, the filament
generally moves against the temperature gradient. In the case when
the influence of heating is positive ($\partial_T f > 0$), as it happens,
for example, when the thermogeneration mechanism is involved, the filament
moves along the temperature gradient.

\subsection{Temperature profile in the moving filament}

For a filament with narrow walls ($\ell_f \ll \ell_T$) the heat
equation (\ref{c_m_2}) is piecewise linear and the temperature
profile $T(\xi)$ can be found analytically (see Appendix).
Consequently, the temperatures $T_L$ and $T_R$ can be presented
explicitly as functions of the filament velocity $v$ and width $W$.
It is convenient to introduce the normalized difference and the sum
of $T_L$ and $T_R$ :
\begin{eqnarray}
\label{DeltaSigma}
\Delta_{LR} \equiv \frac{T_L - T_R}{T^{\star} - T_{\star}},
\qquad
\Sigma_{LR} \equiv
\frac{T_L + T_R - 2 T_{\star}}{T^{\star} - T_{\star}},
\end{eqnarray}
where the temperatures $T^{\star}$ and $T_{\star}$ are stationary uniform
solutions of the heat equation (\ref{RD_T1}) corresponding to uniform on and off
states,
respectively:\cite{Footnote1}
\begin{equation}
\label{T_up_star}
T^{\star} = T_{\rm ext} + {J_{\rm on}(u_{\rm co}) \, u_{\rm co}}/{\gamma},
\end{equation}
and $T_{\star}$ is defined by Eq.\ (\ref{T_low_star}).
Typically $T_{\star} \approx T_{\rm ext}$.

From the solution (\ref{T_profile}) of the heat equation we obtain
\begin{eqnarray}
\label{T_difference}
&&\Delta_{LR} = \frac{1}{\sqrt{1 + \tilde v^2}}
\left(
\tilde v -
\exp \left( - \widetilde W \sqrt {1 + \tilde v^2} \right)
\right. \\ \nonumber
&& \qquad \qquad \times \left. \left[
\tilde v \, \cosh \left(\widetilde W \tilde v \right)
+
\sqrt{1 + \tilde v^2} \,
\sinh \left( \widetilde W \tilde v\right)
\right]
\right), \\
\label{T_average}
&&\Sigma_{LR} = 1 -
\frac{1}{\sqrt{1 + \tilde v^2}}
\exp \left( - \widetilde W \sqrt {1 + \tilde v^2}
\right) \\ \nonumber
&&\qquad \qquad \times \left[
\tilde v \, \sinh \left( \widetilde W \tilde v \right)
+
\sqrt{1 + \tilde v^2} \,
\cosh \left( \widetilde W \tilde v \right)
\right],
\end{eqnarray}
where
\begin{eqnarray}\nonumber
\tilde v = \frac{v}{2 v_T}, \qquad \qquad \widetilde W =
\frac{W}{\ell_T}.
\end{eqnarray}
The dependencies $\Delta_{LR}(v,W)$ and   $\Sigma_{LR} (v,W)$ are
shown in Fig.~3. The temperature difference
is equal to zero for $v = 0$, when the temperature profile is
symmetric, and reaches  maximum at a certain value of $v$ [Fig.~3(a)].
The temperature difference decreases with further increase of $v$ and
vanishes for large velocities ($v \gg W/ \tau_T$) when the filament
moves too fast to heat the semiconductor structure. The average temperature and,
according to Eq.\ (\ref{voltage}), the voltage $u$
monotonically decrease as $v$ increases [Fig.~3(b)]. Hence the onset of
filament motion is always accompanied by a certain voltage drop. Both
$(T_L - T_R)$ and $(T_L + T_R)$ increase with filament width $W$
for a given $v$.

\section{Self-consistent determination of filament velocity}

In the regime of Joule self-heating, the stationary filament motion
with a certain velocity occurs when the moving filament generates the
temperature profile with the temperature difference $(T_L - T_R)$
which is exactly needed to support this motion. Combining Eqs.\
(\ref{velocity_3}),\ (\ref{DeltaSigma}), and (\ref{T_difference}), we
obtain a transcendental equation for the filament velocity $v$:
\begin{eqnarray}
\label{v_final}
&&\frac{v}{v_0} = \Delta_{LR}(v,W), \\
\label{v_0}
&&v_0 \equiv - \frac{T^{\star} - T_{\star}}{{\langle D_a(a) (a_0^{\prime})^2
\rangle}}
\int_{a_{\rm off}}^{a_{\rm on}} \partial_T f \,D_a(a) \, da .
\end{eqnarray}
Here
$v_0$ is the upper limit of the filament velocity which is achieved
for $T_L - T_R = T^{\star} - T_{\star}$.
This velocity characterizes the effect of temperature on
the filament  dynamics. The function $\Delta_{LR}$ is explicitly given by
Eq.\ (\ref{T_difference}). $\Delta_{LR}$ is odd with respect to $v$,
reflecting the symmetry between left and  right directions of
the filament motion.
Hence nontrivial solutions $v^{\star} \ne 0$ of Eq.\ (\ref{v_final}) always come
in
pairs ($v^{\star}, - v^{\star}$). Having this in mind, below we discuss only
nonnegative solutions $v \geq 0$. Note that $v_0 < 0$ for
$\partial_T f > 0$ (positive influence of temperature on vertical
transport), and therefore Eq.\ (\ref{v_0}) has only trivial solution $v = 0$:
filament motion is not possible. We shall focus
on the case $\partial_T f < 0$ when $v_0 > 0$.

It is immediately evident from Fig.~3(a) that
Eq.\ (\ref{v_final}) has either one or two nonnegative roots
depending on $v_0$ and $W$. Solution
$v^{\star} = 0$, corresponding to a steady filament,
exists for all parameter values. Nontrivial solution $v^{\star} > 0$,
that corresponds to the traveling  filament, exists if
\begin{equation}
v_0 \left. \frac{d \Delta_{LR}}{dv}\right|_{v=0}  > 1.
\label{stability}
\end{equation}
When the root $v=0$ is unique, the corresponding stationary filament
is stable because $\Delta_{LR}(v) < v/v_0$ for $v > 0$. This
inequality  means that traveling filament generates a temperature
difference $(T_L - T_R)$ which is not sufficient to support its
motion. The situation changes when the condition (\ref{stability}) is
met: in this case $\Delta_{LR}(v) > v/v_0$ for $v > 0$, and the steady
solution is unstable. It means that the steady filament  looses
stability simultaneously with appearance of the nontrivial solution
$v^{\star} > 0$ which corresponds to traveling filament. Hence,
Eq.\ (\ref{stability}) represents a condition for spontaneous onset of
the filament motion. We discuss the bifurcation from static to
traveling filament in more details in the next section.

Traveling filament is stable because $\Delta_{LR}(v) > v/v_0$ for $v < v^{\star}$
and $\Delta_{LR}(v) < v/v_0$ for $v > v^{\star}$. Indeed, these
inequalities  mean that if $v$ decreases the temperature difference
($T_L - T_R$) increases and hence according to (\ref{velocity_3})
the filament velocity should increase again. In the same way, with increase
of $v$ the difference ($T_L - T_R$) decreases and therefore the
filament slows down.

In Fig.~4(a) we present numerical solutions $v(W)$ of Eq.\ (\ref{v_final})
for different values of the parameter $v_0$.
For a given $v_0$ the motion is possible for filaments whose width exceeds a
certain
threshold $W_{\rm th}(v_0)$ (Fig.~5). With increase of $W$
the filament velocity increases and eventually saturates. We
discuss the analytical approximations of the front velocity in
Sec.\ VII and Sec.\ VIII.

In Fig.~4(b) we present the voltage $u$ in the regime of current filamentation
obtained by substituting numerical solutions
of Eq.\ (\ref{v_final}) into Eqs.\ (\ref{voltage}),\ (\ref{T_average}).
The normalized deviation of $u$ from $u_{\rm co}$
\begin{equation}
\label{u_normal}
\Delta \widetilde u = 2\;
\frac{B^{-1}(u - u_{\rm co}) - (T_{\star} - T_{\rm ext})}{T^{\star}-T_{\star}}
\end{equation}
is shown, where $B$ is determined by Eq.\ (\ref{voltage}).
The voltage drop, clearly visible on curves 2--7, is
associated with the onset of filament motion. With further increase
of $W$ the voltage increases again. Curve 1 is calculated for
$v_0/v_T = 2.1$, which is close to the threshold value $v_0/v_T = 2$,
and practically coincides with the curve for a static filament. Since
the filament width $W$ is directly proportional to the total current
$I$ [see Eq.\ (\ref{width_current})], Fig.~4(b) actually represents the
current-voltage characteristic of the structure with traveling
filament.

\section{Onset of the filament motion}

To analyze the onset of filament motion and propagation of slow
filaments we expand $\Delta_{LR}$
with respect to $v/v_T$ up to the second order:
\begin{eqnarray}
\label{T_difference_slow}
\Delta_{LR}(v,W) &\approx&
\frac{v}{2 v_T}
\left(1 - \frac{1}{8} \left( \frac{v}{v_T} \right)^2 \right)\;A(W),
\\ \nonumber
A(W)&\equiv& \left[ 1 - \left(1 + \frac{W}{\ell_T} \right)\,
\exp \left(- \frac{W}{\ell_T} \right) \right].
\end{eqnarray}
In this case the temperature profile is close to the symmetric
stationary profile given by Eq.\ (\ref{T_profile_steady}).

Substituting Eq.\ (\ref{T_difference_slow}) in Eq.\ (\ref{stability}),
we obtain an explicit condition
for the onset of filament motion
\begin{equation}
\frac{v_0}{2 v_T} A(W) > 1.
\label{stability_1}
\end{equation}
This condition can be further simplified for narrow and wide filaments:
\begin{eqnarray}
\label{stability_2}
\frac{v_0}{4 v_T} \left( \frac{W}{\ell_T} \right)^2 > 1 \; \;
&{\rm for }& \; \; W \ll \ell_T,
\\
\label{stability_2a}
\frac{v_0}{2 e v_T}\left(\frac{W}{\ell_T} + e - 3 \right) > 1 \; \;
&{\rm for}& \; \; W \sim \ell_T,
\\
\label{stability_3}
\frac{v_0}{2 v_T} > 1 \; \;
&{\rm for }& \; \; W \gg \ell_T,
\end{eqnarray}
where $e$ is the natural logarithmic base.
Since $A(W) < 1$, it follows from Eq.\ (\ref{stability_1}) that
regardless to the filament width $W$ the filament motion is
not possible if $v_0 < 2 v_T$. If $v_0 > 2 v_T$, the filaments whose
width $W$ overcomes the threshold $W_{\rm th}$ determined by
Eq.\ (\ref{stability_1}) start to move, whereas smaller
filaments remain steady. The dependence of $W_{\rm th}$ on $v_0/v_T$
is shown in Fig.~5, curve 1. This dependence can be approximated
as
\begin{eqnarray}
\label{W_th_1}
W_{\rm th} \approx 2 \ell_T \sqrt{\frac{v_T}{v_0}} \; \; &{\rm for}&
\; \; W_{\rm th} \ll \ell_T, \\ \label{W_mid}
W_{\rm th} \approx \ell_T \left(2 e \frac{v_T}{v_0}+3-e\right) \; \; &{\rm for}&
\; \; W_{\rm th} \sim \ell_T,
\\ \label{W_th_2}
W_{\rm th} \approx - \ell_T \ln \left(1 - \frac{2 v_T}{v_0} \right)
\; \; &{\rm
for}& \; \; W_{\rm th} \gg \ell_T.
\end{eqnarray}

The bifurcation to traveling filament resembles a supercritical
pitchfork bifurcation:\cite{GUC83} at the bifurcation point
$W = W_{\rm th}$ the static
solution becomes unstable and simultaneously two stable branches
corresponding to the filaments traveling to the left and to the
right appear [Fig.~4(a)]. This bifurcation can be understood in terms of
stability analysis of the current filament performed in
Ref.\ \onlinecite{ALE98} for standard two-component model
(\ref{RD0}),\ (\ref{GC0}): The spectrum of eigenmodes of a static
filament includes a neutral mode $\Psi_T$ with zero eigenvalue
$\lambda_T = 0$ which corresponds to translation. Existence of this
neutral  mode reflects the  translation invariance of a static
filament on a large spatial domain. In the  extended model
(\ref{RD1}),\ (\ref{RD_T1}),\ (\ref{GC1}) the bifurcation to traveling
filament is characterized by symmetry breaking when $\lambda_T$
becomes positive. This becomes possible due to the coupling between
the master equation (\ref{RD1}) and the heat equation (\ref{RD_T1})
when $\partial_T f <0$. Note that for $\partial_T f > 0$ the
eigenvalue of the translation mode $\lambda_T$ becomes negative.
This corresponds to self-pinning of the filament.

\section{Propagation of slow filaments}

Substituting  Eq.\ (\ref{T_difference_slow}) into Eq.\ (\ref{v_final}) we
obtain an explicit equation for the filament  velocity
which is applicable for slow fronts:
\begin{eqnarray}
\label{v_slow}
v(W) = 2 v_T \sqrt{\frac{v_0}{v_T} A(W) - 2},
\end{eqnarray}
where $A(W)$ is defined in Eq.\ (\ref{T_difference_slow}).
The asymptotic value is given by
\begin{equation}
\label{v_slow_asymp}
v \rightarrow 2 v_T \sqrt{\frac{v_0}{v_T} - 2} \qquad {\rm for} \qquad
\frac{W}{\ell_T} \rightarrow  \infty.
\end{equation}
Eq.\ (\ref{v_slow}) approximates $v(W)$ with an accuracy of 10\% up to the
parameter value $v_0/v_T = 3$ [curve 2 on Fig.~4(a)], despite it is
derived for $v \ll v_T.$
This wide range of applicability is due to
particulary smooth behavior of $\Delta_{LR}(v)$ to the left of
its peak value [see Fig.~3(a)].

\section{Propagation of fast filaments}

In the limit $v \gg v_T$ the temperature profile is strongly
asymmetric [see Eq.\ (\ref{T_profile_fast})], and the temperature at the leading
edge is close to $T_{\star}$.
The expression (\ref{T_difference}) can be expanded
with respect to $v_T/v$ :
\begin{eqnarray}
\label{T_difference_fast}
\Delta_{LR}(v,W) = 1 &-& \exp \left(- \frac{W}{v \tau_T} \right)
\\ \nonumber
&-& \left(\frac{v_T}{v}\right)^2 \,
\left[ 2 - \exp \left(- \frac{W}{v \tau_T} \right) \right].
\end{eqnarray}
According to Eq.\ (\ref{eigenvalues}) the
characteristic scale of the temperature profile 
for the fast filament is determined by
$(\lambda^+)^{-1} \approx v \tau_T$ which exceeds $\ell_T$. To keep
the diffusion correction, we expand up to the second order.

Analytical results are available for the case of a narrow filament
($W \ll v \tau_T$) and a wide filament ($W \gg v \tau_T$).
In the first case ($W \ll v \tau_T$) the temperature inside the filament
increases
linearly, and $T_L$
is much smaller than $T^{\star}$. For
$W/v \tau_T \ll 1$, equation (\ref{T_difference_fast})
reduces to
\begin{equation}
\Delta_{LR}(v,W) \approx \frac{W}{\ell_T} \left( \frac{v_T}{v}
\right)
- \left( \frac{v_T}{v} \right)^2.
\label{T_difference_fast_1}
\end{equation}
Substituting (\ref{T_difference_fast_1}) into Eq.\ (\ref{v_final})
we obtain the filament velocity
\begin{equation}
v(W) \approx \sqrt{\frac{W v_0}{\tau_T}} - \frac{\ell_T v_T}{2 W}
\qquad {\rm for } \qquad
W \ll v \tau_T.
\label{velocity_fast_1}
\end{equation}
Note that in Eq.\ (\ref{velocity_fast_1})
the leading term does not depend on the heat diffusion.

In the case of wide filament ($W \gg v \tau_T$) the temperature at
the back edge is close  the maximum value $T^{\star}$.
The front velocity is approximated by
\begin{eqnarray}
\label{velocity_fast_2}
v(W) \approx v_0 - \frac{2 v_T^2}{v_0} -
\exp\left(-\frac{W}{v_0 \tau_T} \right)
\left[v_0  - \frac{2 v_T^2}{v_0}\right]
\\ \nonumber
{\rm for } \qquad
W \gg v \tau_T
\end{eqnarray}
and saturates at
$$
v \rightarrow v_0 - \frac{2v_T^2}{v_0} \qquad
{\rm for} \qquad
\frac{W}{v_0 \tau_T} \rightarrow \infty.
$$

\section{Discussion}

\subsection{Scales hierarchy - which limiting case is relevant?}

The thermal relaxation time $\tau_T$ of the semiconductor structure
can be straightforwardly determined in experiment. Therefore it is
convenient to use it as basic parameter and to express $\ell_T$ and
$v_T$ via $\tau_T$. Depending on the structure width and design,
the effective value of  $\tau_T$ varies from $100 \,$-$200 \;{\rm ns}$
for transistorlike structures of electrostatic discharge protections
devices\cite{POG03} (the structure width
$w \sim 10\;{\rm \mu m}$) to $\sim 10 \,$-$100 \; {\rm ms}$ for power devices
(the structure width $w \sim 100 \,$-$500 \;{\rm \mu m}$) \cite{WAC91}.
It follows from Eq.\
(\ref{thermal_scales}) that parameters $\ell_T$ and $v_T$ are
connected to $\tau_T$ via
\begin{equation}
\ell_T = \sqrt{D_T \tau_T}, \qquad v_T = \sqrt{\frac{D_T}{\tau_T}},
\qquad D_T \equiv \frac{\kappa}{c \rho},
\end{equation}
where the thermal diffusivity $D_T$ depends only on material
parameters. We take $D_T = 0.92 \,{\rm cm^2/s}$
and $D_T = 0.25 \,{\rm cm^2/s}$ for Si and GaAs, respectively.
Consequently,  $\ell_T$ and $v_T$ are of the order of
$100 \; {\rm \mu m}$ and $10^3 \; {\rm cm/s}$, respectively, for small devices
($\tau_T \sim 100 \; {\rm ns}$, $w \sim 10 \; {\rm \mu m}$).
We obtain 1~mm  and 10~cm/s, respectively, for large power
devices ($\tau_T \sim 10\; {\rm ms}$, $w \sim 100 \; {\rm \mu m}$).
Hence  in most devices the filament width $W$ is of the order
of the thermal diffusion length $\ell_T$ or smaller, and
$\ell_T$ is smaller but comparable to the transverse
dimension of the structure $L$. In particular, it implies that in the regime of
self-heating the maximum temperature in the current filament
is much smaller than $T^{\star}$.
For $W \lesssim \ell_f$ and sufficiently far
from the bifurcation point $v_0 = 2 v_T$ the filament velocity is
well described by the formula (\ref{velocity_fast_1}), where the
second term can be neglected.

\subsection{Stimulating the filament motion}

Motion of the filament delocalizes heating of semiconductor structure
and thus is desirable in applications. Generally,  the start of the
filament motion becomes easier with increase of $v_0$ and decrease of
$v_T$ and $\ell_T$ (see Eq.\ (\ref{stability_1}) and Fig.~5, curve
1). However, these parameters cannot be varied independently. Below
we focus on the effect of the structure parameters which enter our
model: the heat transfer coefficient $\gamma$ and  the structure
width $w$.

Curve 2 in Fig.~5 shows the threshold filament width $W_{\rm th}$
normalized by the quantity $(D_T^2 \tau_T / v_0)^{1/3}$
which does not depend on the heat transfer coefficient $ \gamma$.
Taking into account that $v_T/v_0 \sim
\gamma^{3/2}$, we conclude that decrease of $\gamma$ makes the
onset of the filament motion easier. This occurs due to the increase of $v_0$
which, according to Eqs.\ (\ref{T_low_star}),\ (\ref{T_up_star}),\
(\ref{v_0}), scales as
$v_0 \sim \gamma^{-1}$. However, easy start of the filament motion
in structures with inefficient cooling comes at the price of increasing
the temperature in the filament, which makes static filaments
more dangerous.

Curve 3 in Fig.~5 shows $W_{\rm th}$ normalized by the quantity
$(D_T/v_0)^{1/2}$ which does not depend on $w$. Since $v_T/v_0 \sim
w^{-1/2}$, we see that the dependence of $W_{\rm th}$ on $w$ is
nonmonotonic. Physically relevant situation $W \lesssim \ell_T$
corresponds to the left part of the curve 3 where $W_{\rm th}$
increases with $w$. The effect is due to increase of $\ell_T \sim
w^{1/2}$. Thus, for the fixed value of $v_0$, thinner structures are
preferable for the  filament motion.

\subsection{Transient behavior}

Static current filament typically appears  due to the spatial
instability of the uniform state on the middle branch of
current-voltage characteristic (see Fig.~1). This occurs \cite{WAC95}
on the time scale $\tau_a$  which is typically smaller than the
thermal time scale $\tau_T$. First the voltage settles at $u =
u_{\rm co}$, and only then the temperature starts to increase.
According to Eq.\ (\ref{voltage})
the Joule heating is accompanied by increase of  $u$. As soon as the
static filament gets hot, the bifurcation to the
traveling filament occurs, provided that the condition
(\ref{stability}) is satisfied. The onset of filament motion leads to
a certain voltage drop, though $u$ remains larger than
$u_{\rm co}$ [see Eq.\ (\ref{voltage}) and Fig.~4(b)]. The motion can start
before the stationary temperature profile in the static filament is
established, but the nonmonotonic dynamics of $u$ remains
qualitatively the same.

\subsection{Self-motion and self-pinning}

As it has been pointed out in Sec.\ VI, we predict,
instead of self-motion,  self-pinning of the
filament at its initial location due to the Joule self-heating
in the case $\partial_T f > 0 $.
For example, it happens when the
temperature becomes high enough for thermogeneration to set in.
This typically precedes  thermal destruction of the
semiconductor structure due to local overheating.
Eq.\ (\ref{velocity_3}) also suggests an experimental method to distinguish
between positive ($\partial_T f > 0$) and negative ($\partial_T f <
0$) influence of temperature on vertical transport by observing the
filament dynamics in externally applied temperature gradient:
filaments move along and against the temperature gradient for
$\partial_T f > 0$ and $\partial_T f < 0$, respectively.

\subsection{Motion of low-current filaments}

S-shaped current-voltage characteristic exhibits formal duality
between high current density and low current density branches.
Therefore apart from high-current filaments, there are patterns  in
form of low-current filaments: domains of low current density
embedded into on state. \cite{SCH87,ALE98} Such filaments correspond
to the upper part of the filamentary current-voltage characteristic
(Fig.~1), where average current density is close to $J_{\rm on}$. In
the case $\partial_T f < 0$ low-current filaments also undergo the
bifurcation from static to traveling filament. Expressions
(\ref{velocity_3}),\ (\ref{voltage}) remain valid as they are,
whereas in Eqs.\ (\ref{T_difference}),\ (\ref{T_average}) $T^{\star}$
and $T_{\star}$ should be exchanged. The onset of motion results in
the increase, rather than decrease, of the voltage in this case.
Duality between on and off states is broken when vertical transport
is suppressed near the structure boundary  due to a certain process
at the lateral edge of the semiconductor structure, e.g., surface
recombination or surface leak in the $p\,$-$n$ junction. (This effect
can be modelled  by Dirichlet boundary conditions $a=0$ imposed on
the variable $a$ at $x = 0, L_x$.\cite{SCH87}) In this case the
effect of boundaries can not be neglected for average current
densities close to $J_{\rm on}$ even in the limit $L_x \gg \ell_f$.
The current-voltage characteristic, instead of hysteresis, exhibits a
continuous crossover from the filamentary branch to the branch of
quasiuniform high current density states at high current.\cite{SCH87}
Then only high-current filaments are observable.

\subsection{Reaction-diffusion models for traveling spots}

The model (\ref{RD1}),\ (\ref{RD_T1}),\ (\ref{GC1}) belongs to the
same class of three-component reaction-diffusion models as models for
traveling spots in active media discussed in Refs.\
\onlinecite{KRI94,BOD97,NIE97,BOD98,HEM98,BOD02}.
In contrast to the common two-component
activator-inhibitor model,\cite{KER94,MIK94} three-component models
are capable to describe localized traveling patterns not only on a
one-dimensional spatial domain, but also on spatial domains of higher
dimensions. \cite{KRI94,BOD97} The transition from
static to traveling spot takes place with increase of the relaxation
time of the first inhibitor,\cite{KRI94} in the same way as it occurs
in the common two-component model of pulse propagation in excitable
media.\cite{MIK94} The second fast long-range inhibitor plays an
essential role only in the two-dimensional or three-dimensional case:
it prevents lateral spreading of the traveling spot which otherwise
destroys the spatially localized solution and  eventually leads to
development of a spiral wave.\cite{KRI94,BOD97}  

This additional inhibition can be either global\cite{KRI94} or
local.\cite{BOD97} In the first case the inhibitor has the same value
in the whole system. This value depends on the mean value 
of other dynamical variables in the system and is governed by an 
integro-differential equation. This corresponds to the global
coupling of a spatially extended nonlinear system.  
In the second case the additional inhibitor
is governed by a conventional reaction-diffusion
equation. For traveling spots, the difference between these two cases
becomes crucial only when several spots are considered on a two or higher
dimensional domain: the system of several spot is unstable when the
additional inhibition is global, but becomes stable when it is
local.\cite{BOD97} This difference vanishes  when only one spot is 
present, or the spatial domain is one-dimensional.\cite{BOD97}	
Global coupling through the gas phase occurs 
in the surface reactions \cite{MER94,ENG94,MIK03a,MIK03b} 
and can be implemented  as global feedback loop in the light-sensitive
Belousov-Zabotinsky reaction \cite{ZAI70,MIH02}. Implementation of the
respective control loop has allowed to observe localized traveling
patterns in these systems.\cite{MIH02,MIK03b} 

In contrast to the three-component models discussed in
Refs.\ \onlinecite{KRI94,BOD97}, in the model
(\ref{RD1}),\ (\ref{RD_T1}),\ (\ref{GC1}) both inhibitors are
needed for the onset of filament motion already in the one-dimensional
case. Consequently, the bifurcation to traveling filament is
also different. This reflects the fact that we start with a stationary pattern
in a bistable medium with fast global inhibition (voltage $u$).
This global inhibition is due to the external circuit and 
represents an inherent feature of spatio-temporal 
dynamics of a bistable semiconductor: 
For any evolution of the current density pattern
which is accompanied by the variation of the total current, the 
voltage at an external series resistance changes, causing the
variation of the voltage across the device. 
This inhibition is crucial for the existence of current filaments
which become unstable when the global coupling is eliminated by
operating the device in the voltage-controlled
regime.\cite{VOL69,BASS70,ALE98} 
The motion of the filament is induced by the effect of another slow 
diffusive inhibitor (temperature $T$). Similar nonlinear mechanism causes
the motion of current filaments
obtained by numerical simulations in Refs.\ \onlinecite{NIE95,NIE97}.
However, Refs.\ \onlinecite{NIE95,NIE97} focus on a specific type of
multilayer thyristorlike structures. The model of these devices\cite{NIE92}
assumes that the second inhibitor is a certain internal voltage,
not a temperature.

It is worth to mention, that self-heating may also trigger
temporal relaxation-type oscillations of a current-density
filament. This effect has been observed
in a reversely biased  $p\,$-$i\,$-$n$ diode and
is explained  by a similar three-component
reaction-diffusion model.\cite{DAT97}

\section{Summary}

Joule self-heating of a current density filament in a bistable
semiconductor structure  might result in the onset of lateral
motion.  This occurs when increase of temperature has negative impact
on the vertical (cathode-anode) transport. Such negative feedback
takes place when bistability of semiconductor structure is related to
the avalanche impact ionization, because the impact ionization rate
decreases with temperature.\cite{SZE}  Examples of such devices are avalanche
transistors,\cite{OSI73} reversely biased $p\,$-$i\,$-$n$ diodes in the regime
of avalanche injection, electrostatic discharge protection devices
operated in the avalanche regime,\cite{POG02} multilayer
thyristorlike structures.\cite{GOR02,NIE92} Traveling
filaments can be consistently described by a generic nonlinear model
(\ref{RD1}),\ (\ref{RD_T1}),\ (\ref{GC1}).

Generally,  filaments move against the temperature gradient with a
velocity proportional to the temperature difference at the filament
edges [Eq.\ (\ref{velocity_3})]. In the regime of Joule self-heating
the strength of coupling between the master equation (\ref{RD1}),
which controls the current density dynamics, and the heat equation
(\ref{RD_T1}) can be characterized by single parameter $v_0$ [Eq.\
(\ref{v_0})], which has a dimension of velocity. The filament
velocity $v$ is determined by the transcendental equation
(\ref{v_final}). The condition for the spontaneous onset of filament
motion (\ref{stability}) depends on the ratio of $v_0$ and the
thermal velocity $v_T$, and on the ratio of the filament width $W$
and the thermal diffusion length $\ell_f$. Filament motion is never
possible for $v_0 < 2 v_T$. For $v_0 > 2 v_T$, static filaments whose
width $W$ exceeds a certain threshold $W_{\rm th}$  [Eq.\
(\ref{stability_1}), curve 1 in Fig.~5] become unstable and start to
travel. The filament velocity $v$ and the voltage on the structure
$u$ are shown in Fig.~4. For most semiconductor structures $W
\lesssim \ell_T$, and sufficiently far from the bifurcation point
$v_0 = 2 v_T$ the filament velocity can be approximated by a
truncated version of Eq.\ (\ref{velocity_fast_1})
$$
v \approx \sqrt{\frac{W v_0}{\tau_T}}.
$$
Heating of the static filament is accompanied by an increase of the
voltage $u$ [Eq.\ (\ref{voltage})]. With onset of the  filament motion the
voltage drops, but remains higher than the voltage $u_{\rm co}$
[Fig.~4(b)].

Our analytical theory does not cover narrow filaments, when the flat
top of current density profile does not exist or the filament width
$W$ is comparable to the width of the filament wall $\ell_f$,
as well as cylindrical filaments which appear when lateral dimensions
of semiconductor structure $L_x$ and $L_y$ are comparable.
However, the bifurcation from static to traveling filament and
the mechanism of filament motion remain qualitatively the same in these
cases.

\acknowledgments

Stimulating discussions with S. Bychkhin, M. Denison and D. Pogany
and the discussion of the results
are gratefully acknowledged. The author thanks  A. Mikhailov
and V. Zykov for helpful discussions of the results.
The work was supported by the Alexander von Humboldt Foundation.

\appendix*

\section{Solutions of the heat equation}


Assuming that the front walls are thin, we present Eq.\ (\ref{c_m_2})
as (see Fig.~2):
\begin{eqnarray}
\label{piesewiseheat}
\ell_T^2 \, T^{\prime \prime} + v \, \tau_T \, T^{\prime}
+ (T^{\star}- T) &=& 0 \\ \nonumber
&{\rm for}& \;
- \frac{W}{2} < \xi < \frac{W}{2}, \\ \nonumber
\ell_T^2 \, T^{\prime \prime} + v \, \tau_T \, T^{\prime}
+ (T_{\star}- T) &=& 0 \\ \nonumber
&{\rm for}& \;
\xi < - \frac{W}{2} \;  {\rm and} \;  \xi > \frac{W}{2}.
\end{eqnarray}
Here the middle of the filament is at $\xi = 0$.
$T_{\star}$ and $T^{\star}$ are defined by
Eqs.\ (\ref{T_low_star}),\ (\ref{T_up_star}). The
solution of this piecewise linear equation is given by
\begin{eqnarray}
\label{T_profile}
T (\xi) &=& T_{\star}
- (T^{\star} - T_{\star}) \\ \nonumber
&\times&
\frac{2 \lambda^{+}}{\lambda^{+} - \lambda^{-}}
\sinh \left( \frac{\lambda^{-} W}{2}\right) \,
\exp(\lambda^{-} \xi) \\
\nonumber
&{\rm for}& \qquad
\xi > \frac{W}{2}, \\
\nonumber
T(\xi) &=& T^{\star} -
\frac{T^{\star} - T_{\star}}{\lambda^{+} - \lambda^{-}}
\left(
\lambda^{+} \exp \left[ \lambda^{-} \left( \xi + \frac{W}{2} \right) \right]
\right.
\\ \nonumber
&-&
\left.
\lambda^{-} \exp \left[ \lambda^{+} \left( \xi - \frac{W}{2} \right) \right]
\right)  \\ \nonumber
&{\rm for}& \qquad
\frac{W}{2} < \xi < \frac{W}{2}, \\
\nonumber
T(\xi) &=& T_{\star}
- (T^{\star} - T_{\star}) \\ \nonumber
&\times& \frac{2 \lambda^{-}}{\lambda^{+} - \lambda^{-}}
\sinh \left( \frac{\lambda^{+} W}{2}\right) \,
\exp(\lambda^{+} \xi) \\ \nonumber
&{\rm for}& \qquad \xi < -\frac{W}{2},
\end{eqnarray}
where $\lambda^{+}$ and $\lambda^{-}$ are eigenvalues of
Eq.\ (\ref{piesewiseheat}):
\begin{eqnarray}
\label{eigenvalues}
\lambda^{+} &=&
\frac{1}{\ell_T}
\left[
\sqrt{1 + \left( \frac{v}{2 v_T} \right) ^2}
- \frac{v}{2 v_T}
\right],\\ \nonumber
\lambda^{-} &=&
- \frac{1}{\ell_T}
\left[
\sqrt{1 + \left( \frac{v}{2 v_T} \right) ^2}
+ \frac{v}{2 v_T}
\right].
\end{eqnarray}
Eqs.\ (\ref{T_profile}) allow to determine the temperature
difference $(T_L + T_R)$ and the mean temperature
$(T_L + T_R)/2$
[Eqs.\ (\ref{DeltaSigma}),\ (\ref{T_difference}),\ (\ref{T_average})]
which are  needed for self-consistent calculation of the front
velocity $v$ and the voltage $u$.

{\it Steady temperature profile.}---
For $v = 0$ we get $\lambda^{+} = - \lambda^{-}=1/\ell_T$
and the corresponding symmetric temperature profile is given by
\begin{eqnarray}
\nonumber
T (\xi) &=& T_{\star}
+ (T^{\star} - T_{\star}) \sinh \left( \frac{W}{2 \ell_T} \right)
\exp \left( - \frac{\xi}{\ell_T} \right) \\ \label{T_profile_steady}
&{\rm for }& \qquad \xi > \frac{W}{2}, \\
\nonumber
T(\xi) &=& T^{\star} -
(T^{\star} - T_{\star})
\exp \left( - \frac{W}{2 \ell_T} \right)
\cosh \left( \frac{\xi}{\ell_T}  \right)
\\ \nonumber
&{\rm for }& \qquad
\frac{W}{2} < \xi < \frac{W}{2}, \\
\nonumber
T(\xi) &=&
T_{\star}
+ (T^{\star} - T_{\star}) \sinh \left(\frac{W}{2 \ell_T} \right)
\exp \left( \frac{\xi}{\ell_T} \right) \\ \nonumber
&{\rm for }& \qquad \xi < -\frac{W}{2}.
\end{eqnarray}
{\it Temperature profile in the fast filament.}---
For $v \gg v_T$ we get $\lambda^{+} = (v \tau_T)^{-1}$,
$\lambda^{-} = - \infty$. The corresponding profile is strongly
asymmetric:
\begin{eqnarray}
\label{T_profile_fast}
T (\xi) &=& T_{\star} \qquad {\rm for } \qquad
\xi > \frac{W}{2}, \\
\nonumber
T(\xi) &=& T^{\star} -
(T^{\star} - T_{\star})
\exp \left( \frac{2 \xi - W}{2 v \tau_T} \right)
\\ \nonumber
&{\rm for }& \qquad
-\frac{W}{2} < \xi < \frac{W}{2}, \\
\nonumber
T(\xi) &=&
T_{\star} + 2(T^{\star} - T_{\star}) \sinh \left(\frac{W}{2 v \tau_T}\right)
\exp \left(\frac{\xi}{v \tau_T } \right) \\ \nonumber
&{\rm for }&
\qquad
\xi < -\frac{W}{2}.
\end{eqnarray}

\nopagebreak


\begin{figure*}[hp]
\begin{center}
\includegraphics*[height=8.6cm,width=6.02cm,angle=270]{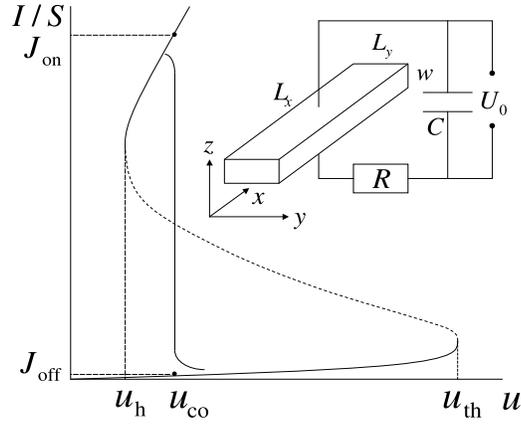}
\end{center}
\caption{
Current-voltage characteristic of a bistable structure.
The average current density $I/S$ is shown,
where $S = L_x L_y$ is the cross-section of the structure.
Unstable middle branch with negative differential conductance
is depicted by the dashed line.  The hold and threshold voltages
are denoted as $u_{\rm h}$
and $u_{\rm th}$, respectively. The vertical branch at $u=u_{\rm co}$
corresponds to a static filament.
The inset shows a sketch of a bistable semiconductor structure operated
in the external circuit with load resistance $R$, capacitance $C$, and
bias $U_0$.
}
\label{CVC}
\end{figure*}

\begin{figure*}[hp]
\begin{center}
\includegraphics*[height=12.9cm,width=9.03cm,angle=270]{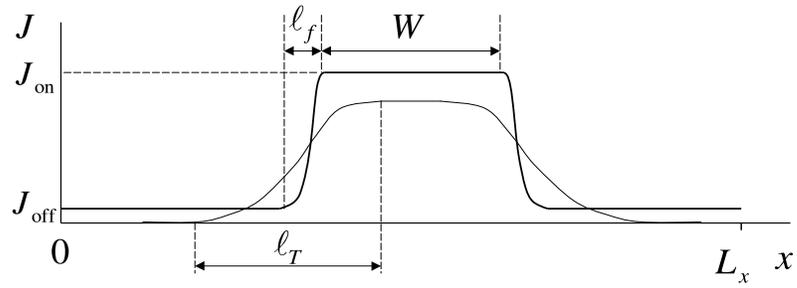}
\end{center}
\caption {Current density profile in a filament (thick line).
Thin line denotes the temperature profile in the static filament. }
\label{profiles}
\end{figure*}

\begin{figure*}[hp]
\begin{center}
\includegraphics*[height=8.6cm,width=6.02cm,angle=270]{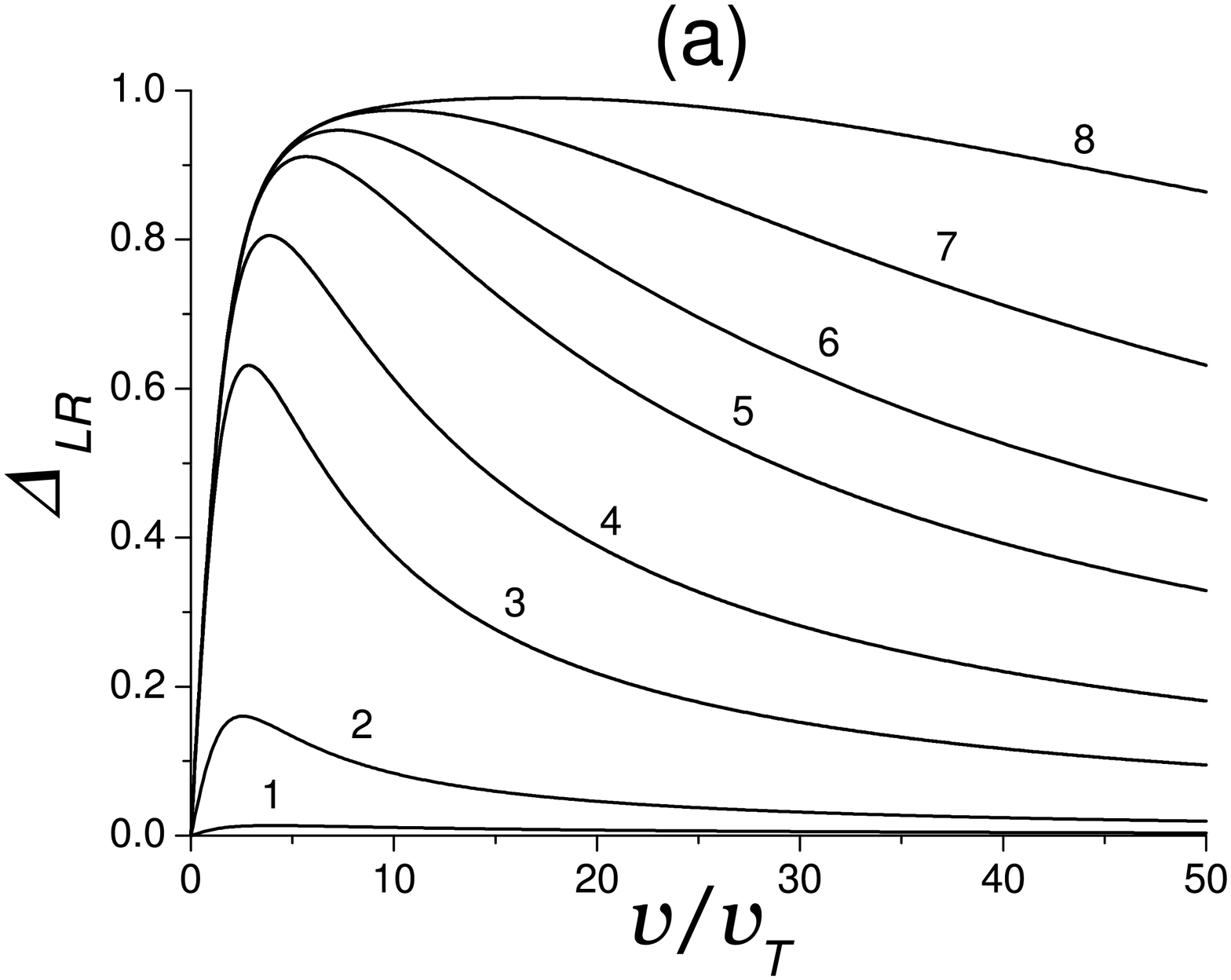}
\includegraphics*[height=8.6cm,width=6.02cm,angle=270]{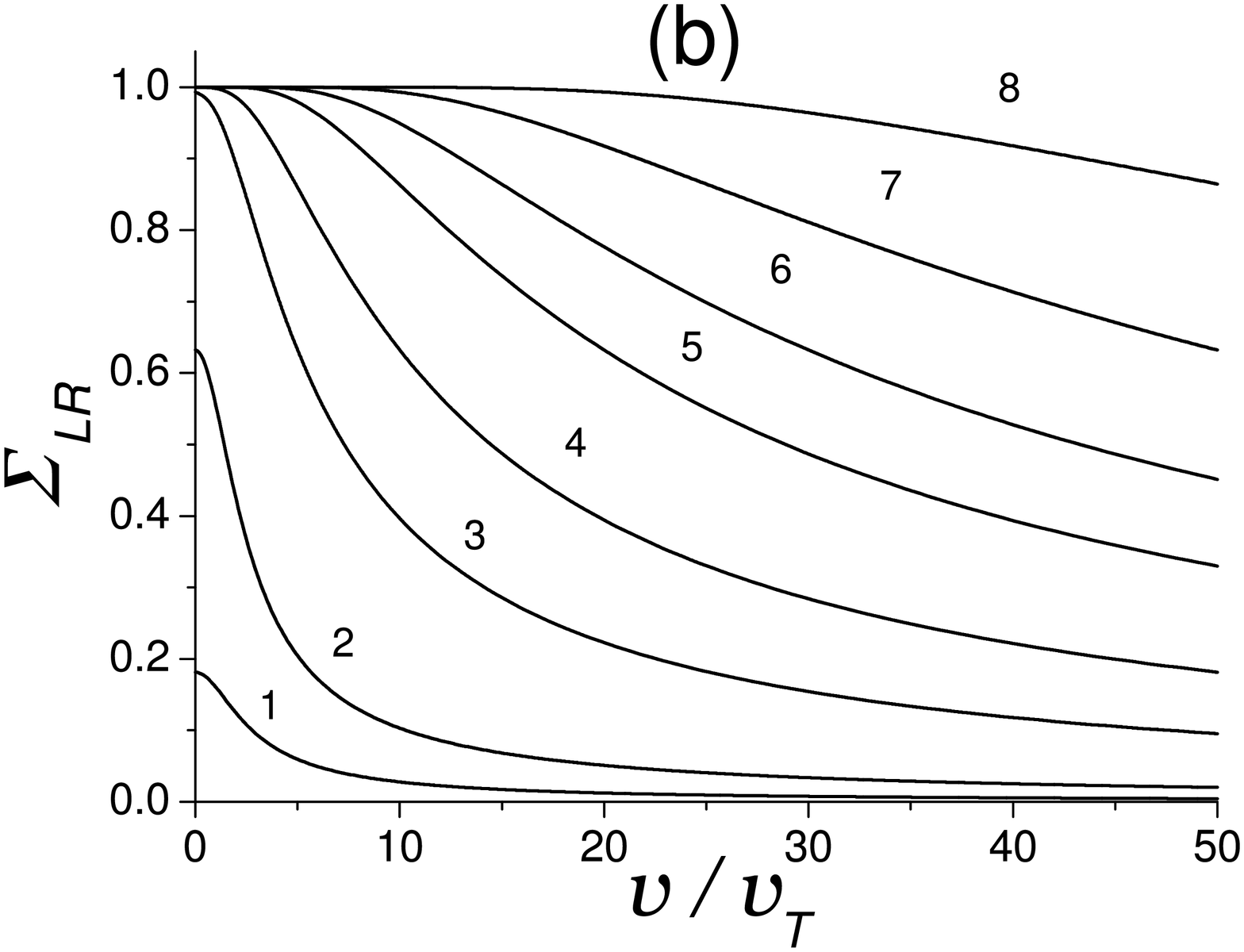}
\end{center}
\caption
{The normalized difference $\Delta_{LR}$~~(a) and the sum $\Sigma_{LR}$~~(b)
of the temperatures in the filament walls $T_L$ and $T_R$ as functions
of the filament velocity $v$ for different filament widths $W$.
$\Delta_{LR}$ and $\Sigma_{LR}$ are defined by
Eqs.\ (\ref{DeltaSigma}),\ (\ref{T_difference}),\ (\ref{T_average}).
Curves 1 to 8 correspond to
$W/\ell_T = 0.2, 1, 5, 10, 20, 30, 50, 100$,
respectively.}
\label{voltage1}
\end{figure*}

\begin{figure*}[hp]
\begin{center}
\includegraphics*[height=8.6cm,width=6.02cm,angle=270]{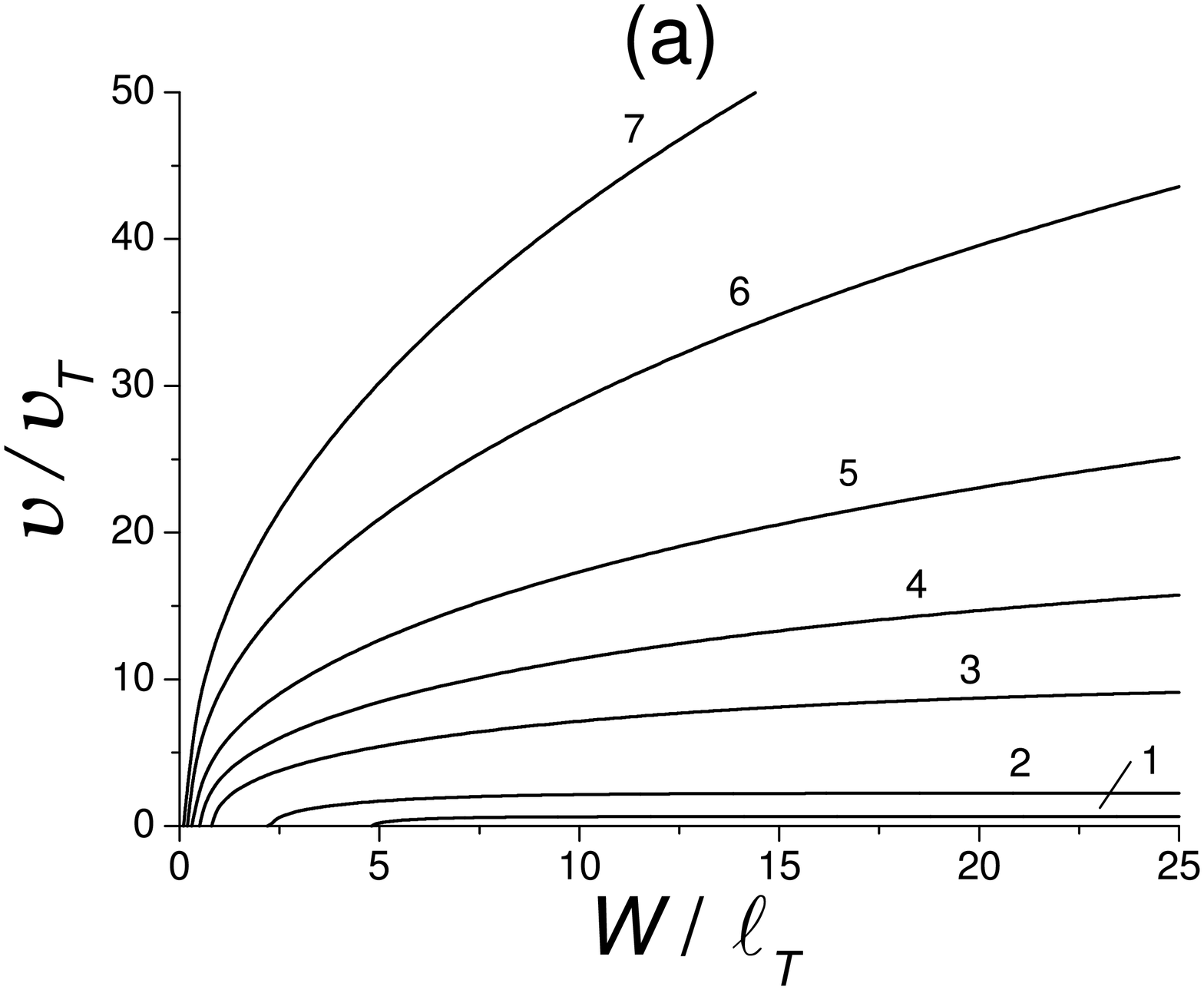}
\includegraphics*[height=8.6cm,width=6.02cm,angle=270]{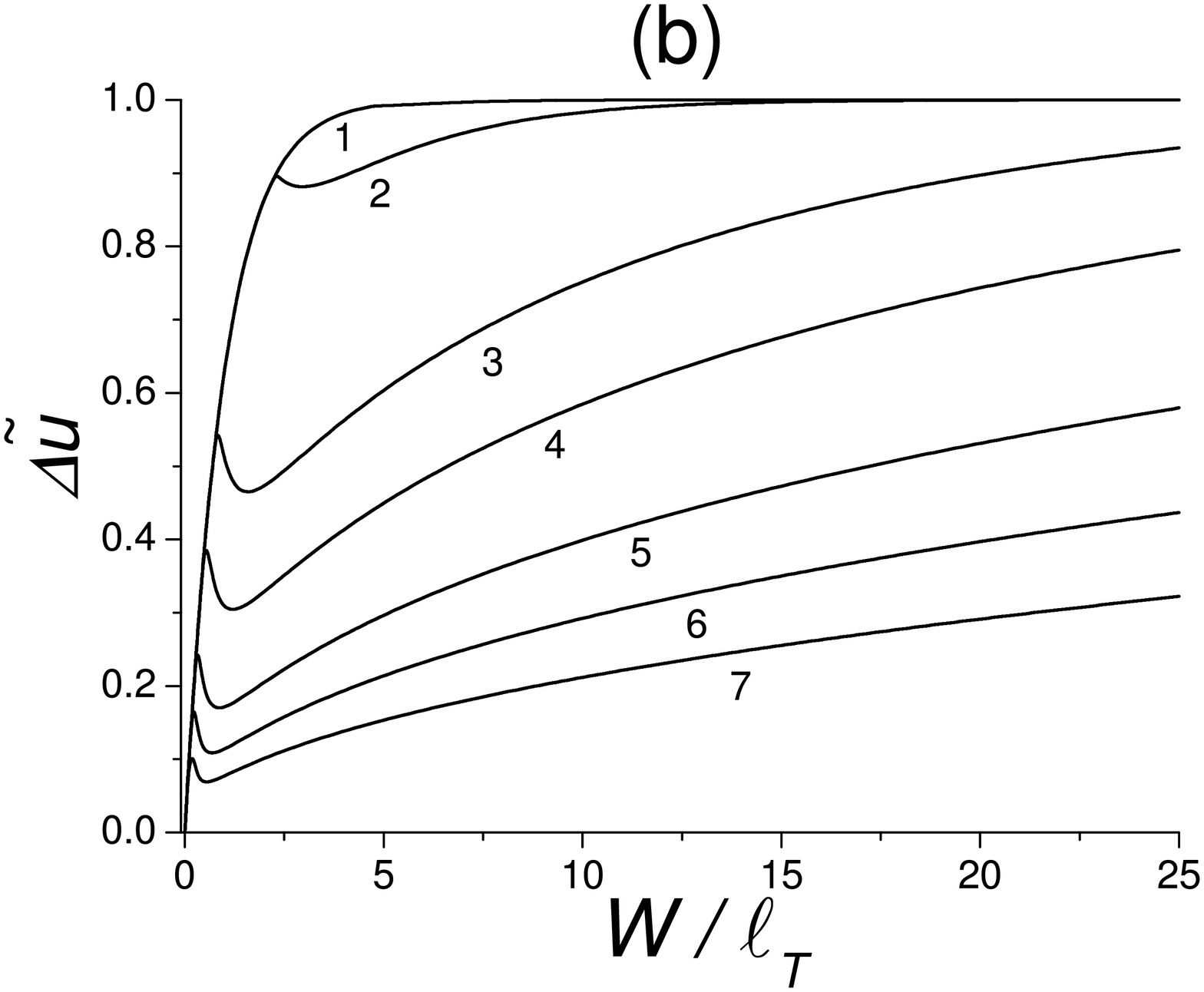}
\end{center}
\caption
{The filament velocity $v$~~(a) and normalized voltage
on the structure
$\Delta \tilde u$~~(b)
as functions of the filament width $W$ for different values of
the parameter $v_0$. $\Delta \tilde u$ is defined by Eq.\ (\ref{u_normal}).
Curves 1 to 7 correspond to
$v_0 = 2.1, 3, 10, 20, 50, 100, 200$,
respectively. Peaks on the voltage curves are related to onset
of the filament motion.
}
\label{vel_vol}
\end{figure*}

\begin{figure*}[hp]
\begin{center}
\includegraphics*[height=8.6cm,width=6.02cm,angle=270]{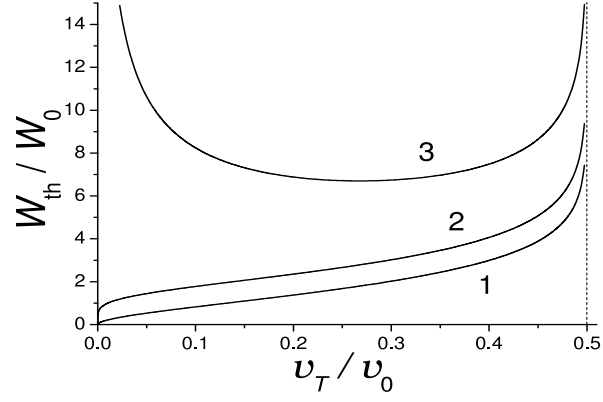}
\end{center}
\caption {Normalized threshold filament width $W_{\rm th}$
corresponding to  onset of the filament motion as a function of $v_T
/ v_0$. Curve 1 shows  $W_{\rm th}$ normalized to the thermal
diffusion length: $W_0 = \ell_T$. Curve 2 shows $W_{\rm th}$
normalized to the quantity $W_0 = (D_T^2 \tau_T/v_0)^{1/3}$, which
does not depend on the heat transfer coefficient $\gamma$. Curve 3
shows  $W_{\rm th}$ normalized to the quantity $W_0 =
\sqrt{D_T/v_0}$, which does not depend on the structure width $w$. }
\label{width}
\end{figure*}


\begin{thebibliography}{99}
\bibitem{Footnote1}
Here we negelect dependence of $T^{\star}$ and $T_{\star}$
on $u$, assuming $u = u_{\rm co}$.
This does not change the results.
\end{thebibliography}
\end{document}